\documentclass[aps,prl,twocolumn,superscriptaddress]{revtex4}

\usepackage{mathptmx}
\usepackage{subfigure}
\usepackage{dcolumn}
\usepackage{amsmath,amssymb}
\usepackage{bm}
\usepackage{color}
\usepackage{latexsym}
\usepackage[english]{babel}
\usepackage{times}

\usepackage{psfrag,graphicx} 
\usepackage{epsf}  
\usepackage{amsfonts}
\usepackage{natbib}
\usepackage{epstopdf}
\usepackage{appendix}
\usepackage[colorlinks=true,citecolor=blue,linkcolor=blue]{hyperref}

\begin{document}

\title{Quasiparticle lifetime in ultracold fermionic mixtures with density and mass imbalance}

\author{Zhihao Lan}
\email{z.lan@soton.ac.uk}
\affiliation{School of Mathematics, University of Southampton, Highfield, Southampton, SO17 1BJ, United Kingdom}

\author{Georg M. Bruun}
\affiliation{Department of Physics and Astronomy, University of Aarhus, Ny Munkegade, DK-8000 Aarhus C, Denmark}

\author{Carlos Lobo}
\affiliation{School of Mathematics, University of Southampton, Highfield, Southampton, SO17 1BJ, United Kingdom}

\date{\today}

\begin{abstract}
We show that   atomic Fermi mixtures  with density and mass imbalance exhibit a 
 rich diversity of scaling laws for the quasiparticle decay rate  beyond the  quadratic energy and temperature dependence  of conventional 
 Fermi liquids. For certain densities and mass ratios, the decay rate is linear whereas in other cases it exhibits a plateau. Remarkably,
  this plateau extends from the deeply degenerate to the high temperature classical regime of the light species. 
 Many of these scaling laws are analogous to what is found in very different systems including dirty metals, liquid metals, and high temperature plasmas.
 The Fermi mixtures can in this sense  span a whole range of seemingly diverse and separate physical systems. Our results are derived in the weakly interacting 
 limit making them quantitatively reliable. 
  The different regimes can be detected with radio-frequency  spectroscopy. 
\end{abstract}


\maketitle

{\it Introduction}.---
There are many systems in Nature composed of two different types of fermions with either different concentrations or different masses such as electron-ion plasmas, liquid metals, spin-polarised conductors and certain cases of nuclear matter. The temperature in these systems usually lies in a characteristic range with respect to the Fermi temperatures of the two components. For example, in spin-polarised conductors, both spin states are fully degenerate while in hot electron-ion plasmas both are fully classical. An interesting class of systems appears when one of the components is degenerate while the other is classical. This occurs for example in spin-polarised Fermi gases when the minority component is classical \cite{zwierlein_polaron, polaron_rudi}, liquid metals and in warm dense matter, a type of plasma where the ions are classical but the electrons are close to degeneracy \cite{wdm}. 
We show in this Letter that two-component Fermi mixtures of ultracold atoms provide the possibility for the first time to access all of these disparate regimes within a single experimental system. For example, mixtures of $^6$Li and $^{40}$K have already been created \cite{polaron_rudi} 
where the $^6$Li atoms are degenerate but where the $^{40}$K atoms can be either above or below their Fermi temperature. 
In these mixtures, we demonstrate that  various regimes of temperature and concentration difference  give rise to  a variety of  energy and temperature scalings
 of  the lifetime of the majority quasiparticles, that are analogous to several of the  very different Fermi mixtures in Nature. 
Contrary to plasmas or the electron gas which in general are extremely complex to describe,
ultracold atomic gases interact via a short range interaction which is accurately  characterised by the scattering length, and
the effects discussed in this paper are all realised in the weak coupling regime. 
  This means that the quasiparticle lifetime is described by a compact and reliable expression involving the Lindhard function, which nevertheless 
  contains all the different physics in various in limits and the interpolations between them.

The reason for the appearance of these scaling laws is that in a system with two types of atom (which we will denote by ``$\uparrow$" and ``$\downarrow$") of different masses and/or densities, the Fermi energies 
of the two species are in general unequal leading to the existence of two energy (and temperature) scales. We will assume here that 
$\epsilon_{F\downarrow}\ll\epsilon_{F\uparrow}$. The difference can be due to a large density imbalance $n_\downarrow\ll n_\uparrow$ with equal mass or 
large mass imbalance $m_\downarrow\gg m_\uparrow$ with equal density. 
  This results in an intermediate energy regime 
  $\epsilon_{F \downarrow }\ll \epsilon \ll \epsilon_{F \uparrow }$  which increases in size with mass or density imbalance. 
While the quadratic behaviour of Fermi liquids is due to the effect of Pauli blocking of both species \cite{Landau},  the key property of this intermediate regime is that the Pauli blocking of the $\downarrow$ atoms  is unimportant so that the phase space for interspecies scattering depends only on the  $\uparrow$ atoms. We will show that, for the $n_\downarrow\ll n_\uparrow$, equal mass mixture, the Pauli blocking of $\uparrow$ atoms  results in a linear decay rate in temperature while for a $m_\downarrow\gg m_\uparrow$, equal density mixture, energy conservation restricts the phase space for scattering of the $\uparrow$ atoms to a narrow region around the Fermi surface, making Pauli blocking of the $\uparrow$ atoms irrelevant, leading to a temperature-independent decay rate. The effects on the lifetime due to modification of Pauli blocking in spin polarized Fermi systems, were known in spin polarized liquid $^3$He and ferromagnetic metals ~\cite{jltp, mineev, fm}. For example they are responsible for the observed
zero temperature damping of the transverse spin dynamics in spin polarized $^3$He~\cite{jltp, mineev}. 
Experimentally, Fermi mixtures with density imbalance~\cite{ketterle, salomon_nature} or mass imbalance ~\cite{Fermi_Fermi1, Fermi_Fermi3, opticallattice} have been 
 created which raises the prospect of observing the effects we discuss here in the near future using for instance radio-frequency (RF)  spectroscopy.

{\it Model}.--- We consider a homogeneous gas of two species of fermions denoted $\sigma=\uparrow,\downarrow$ with masses $m_\uparrow\leq m_\downarrow$ 
and densities $n_\uparrow\ge n_\downarrow$, from which we define the Fermi momenta $k_{F\sigma} \equiv (6 \pi^2 n_{\sigma})^{1/3}$.
The key quantity we study in this paper is the decay rate $1/\tau_p$ of the $\uparrow$ quasiparticles with momentum ${\mathbf p}$
and energy $\epsilon_{p\uparrow}$.
To lowest order in the scattering processes, the decay rate can be written as~\cite{BaymPethick} 
\begin{eqnarray}\label{lifetime}
\frac1{\tau_p}=2\pi U^2\sum_{\bf k} \sum_{\bf q} \delta( \epsilon_{\bf p\uparrow }+\epsilon_{ \bf k \downarrow}-\epsilon_{{{\bf p-q}\uparrow }}
-\epsilon_{ {\bf k+q}\downarrow}) \nonumber \\
 \times [n_{\bf k\downarrow } (1-n_{ {\bf k+q}\downarrow} ) (1-n_{{\bf p-q}\uparrow } )+(1-n_{ \bf k\downarrow} )n_{{\bf k+q}\downarrow }
  n_{{\bf p-q}\uparrow }] 
\label{lmSigma}
\end{eqnarray}
where  $n_{\bf k\sigma} =(e^{\beta \xi_{\bf k\sigma}}+1)^{-1}$ is the Fermi function.
We have defined $\xi_{\bf p\sigma}=\epsilon_{\bf p\sigma}-\mu_\sigma$  with $\mu_\sigma$
the chemical potential, and $\beta=1/T$ with $T$ the temperature (we set $k_B=\hbar=1$).
The energy is    $\epsilon_{p\sigma}=  p^2/2m_\sigma$ where $m_\sigma$ is the effective mass.
The parameter $U$ is the effective interaction between the $\uparrow$ and the $\downarrow$ atoms, and we have 
neglected interactions between identical atoms for simplicity. In the strong coupling regime, one can extract the value of the effective mass and $U$ from Monte Carlo, variational and thermodynamic arguments~\cite{bruun,polaron_theory, MassignanEPJD}. We will mostly work in the weak coupling regime, where we have $U=2 \pi a/m_r$ and where $m_r=m_\uparrow m_\downarrow/(m_\uparrow+m_\downarrow)$ is the reduced mass and $a$ the scattering length for the interaction between the two atom species \cite{note2}.

For analytic investigation, it is convenient to rewrite (\ref {lmSigma}) in terms of the imaginary part of the Lindhard  
function of the $\downarrow$ atoms given by \cite{book_vignale}
 \begin{equation}
  {\rm Im} \chi_\downarrow(q,\omega)=-\pi\int \frac{d^3k}{(2\pi)^3}(n_{\bf k\downarrow} -n_{ \bf k+q \downarrow} ) 
  \delta(  \omega-\epsilon_{\bf k+q\downarrow}+\epsilon_{\bf k\downarrow}). 
 \end{equation}
To do this, we first recast the $\delta$ function in (\ref {lmSigma})  in the form 
$\delta( \epsilon_{\bf p\uparrow}+\epsilon_{\bf k\downarrow}-\epsilon_{\bf p-q\uparrow}-\epsilon_{\bf k+q\downarrow}) =
\int_{-\infty}^{+\infty} d\omega \delta(\omega-\epsilon_{\bf p\uparrow}+\epsilon_{\bf p-q\uparrow}) \delta( \omega-\epsilon_{\bf k+q\downarrow}+\epsilon_{\bf k\downarrow})$.
We also use the Fermi function identities $n_{\bf k\downarrow} (1-n_{\bf k+q\downarrow} ) =(n_{\bf k\downarrow} -n_{\bf k+q\downarrow} )/(1-e^{-\beta\omega})$
 and $(1-n_{\bf k\downarrow}) n_{\bf k+q\downarrow} =(n_{\bf k\downarrow} -n_{\bf k+q\downarrow} )/(e^{\beta\omega}-1)$, with 
 $\omega \equiv \epsilon_{\bf k+q\downarrow}-\epsilon_{\bf k\downarrow}$. 
 Finally, the angular integral over ${\bf q}$ is $\int\Omega_{q} \delta(\omega-\epsilon_{\bf p\uparrow}+\epsilon_{\bf p-q\uparrow})=
 2\pi m_{\uparrow}/pq $ with $ -pq/m_{\uparrow}-q^2/2m_{\uparrow}\leq\omega\leq pq/m_{\uparrow}-q^2/2m_{\uparrow}$
  and we obtain
\begin{equation}\label {ImSig}
\frac{1}{\tau_p}=-\frac{ m_{\uparrow}|U|^2}{2\pi^2 p}
\int_0^\infty dq   q\int_{\omega_-}^ {\omega_+}d\omega{\rm Im} \chi_{\downarrow}(q,\omega) F(\omega, \epsilon_{\bf p\uparrow},\mu_\uparrow )
\end{equation}
 where  $ F(\omega, \epsilon_{\bf p\uparrow},\mu_\uparrow )= (1+e^ {\beta (\omega-\xi_{\bf p\uparrow})})^{-1}(1-e^{-\beta\omega})^{-1}+
 (1+e^ {-\beta (\omega-\xi_{\bf p\uparrow})})^{-1}(e^{\beta \omega}-1)^{-1} $, and
$\omega_\pm(q)=\pm pq/m_\uparrow-q^2/2m_\uparrow$.

Before proceeding, 
let us briefly discuss the difference between the situation considered here and the usual quadratic Fermi liquid behaviour. 
In a conventional Fermi liquid, the low energy condition 
$\xi_p=\epsilon_p-\mu\ll \epsilon_F$ and  $T\ll\epsilon_F$ ensures that one can use ${\rm Im} \chi_{\downarrow}(q,\omega) \propto \omega/q$ 
in (\ref{ImSig}) leading to quadratic scaling of the decay rate with energy and temperature. 
Here, the Lindhard function of the $\downarrow$ atoms has a  different behaviour in the intermediate regime, which will result in different power laws.

{\it Zero temperature}.--- In the following we will take $\xi_{p \uparrow } \geq 0$ without loss of generality since $\tau_p(\xi)$ 
is an even function for $|\xi| \ll \epsilon_{F \uparrow }$. In this case, the back-scattering term [the second term in  (\ref{ImSig})] vanishes at $T=0$. The integration region 
 in  (\ref{ImSig}) is determined by three conditions: from the Bose factors we have $0\le\omega\le\xi_{p \uparrow }$; 
$\omega \le \omega_+(q)$ and $0\le q\le 2p$.
Using the  $T=0$ expression for the 
Lindhard function~\cite{book_vignale} , 
$ \text{lm}\chi_{\downarrow}(q,\omega)=-m_{\downarrow}^2\epsilon_{F \downarrow }/4\pi q[\Theta(1-v_{-}^2)(1-v_{-}^2)-\Theta(1-v_{+}^2)(1-v_{+}^2)]$ 
with $v_{\pm }=m_{\downarrow} \omega/qk_{F \downarrow } \pm q/2k_{F \downarrow }$,
we obtain 
\begin{gather}
\frac1{\tau_p}=
\frac{|U|^2 m_{\uparrow}m_{\downarrow}^2\epsilon_{F\downarrow}} {8\pi^3 p}
\int_0^{2p} dq \int_0^{\xi_{p\uparrow}, \hspace{1mm} \omega_{+}} d\omega   \nonumber \\ 
\times [\Theta(1-v_{-}^2)(1-v_{-}^2)-\Theta(1-v_{+}^2)(1-v_{+}^2)].
\end{gather}

(i) The low energy regime  $\xi_{\bf p \uparrow} \ll \epsilon_{F \downarrow } \ll \epsilon_{F \uparrow } $. 
 When the energy is small compared to 
both Fermi energies, we can use ${\rm Im} \chi_{\downarrow}(q,\omega)=-m_\downarrow^2\omega/4\pi q$ which gives
\begin{equation}
\frac1{\tau_p}=\frac{|U|^2 m_{\uparrow}m_{\downarrow}^2 k_{F\downarrow}} {8\pi^3 p}\xi_{p\uparrow }^2.
\label{exc1}
\end{equation}
This is the usual quadratic dependence of the decay rate  on excitation energy characteristic of a conventional Fermi liquid. 
Indeed, when $m_\downarrow=m_\uparrow$ we recover the well-known expression for the damping rate of a quasiparticle at 
$T=0$~\cite{book_vignale}.

(ii) The intermediate regime {\it $\epsilon_{F\downarrow} <\xi_{\bf p \uparrow} < \gamma\epsilon_{F\downarrow} \ll\epsilon_{F\uparrow} $} with  $\gamma=4(k_{F\uparrow} /k_{F\downarrow} -1)(k_{F\uparrow} /k_{F\downarrow} +m_{\uparrow} /m_{\downarrow} )/(1+m_{\uparrow} /m_{\downarrow} )^2$ where $\gamma \epsilon_{F\downarrow}$ is defined as the $\omega$ coordinate of the intersection of the $\omega_+$ and $\nu_-=-1$ curves  (for this region to exist, apart from the main effect of mass imbalance, we must have $\gamma>1$ which also sets a condition on the density imbalance) .  We obtain
\begin{equation}
\frac1{\tau_p}= \frac{|U|^2 m_{\uparrow}m_{\downarrow}^2 n_{F\downarrow}} {2\pi p } (\xi_{p\uparrow}-\frac{2}{5}\epsilon_{F\downarrow}).
\label{exc2}
\end{equation}

\begin{figure}[!hbp]
\centering
 \includegraphics[width=0.8\columnwidth]{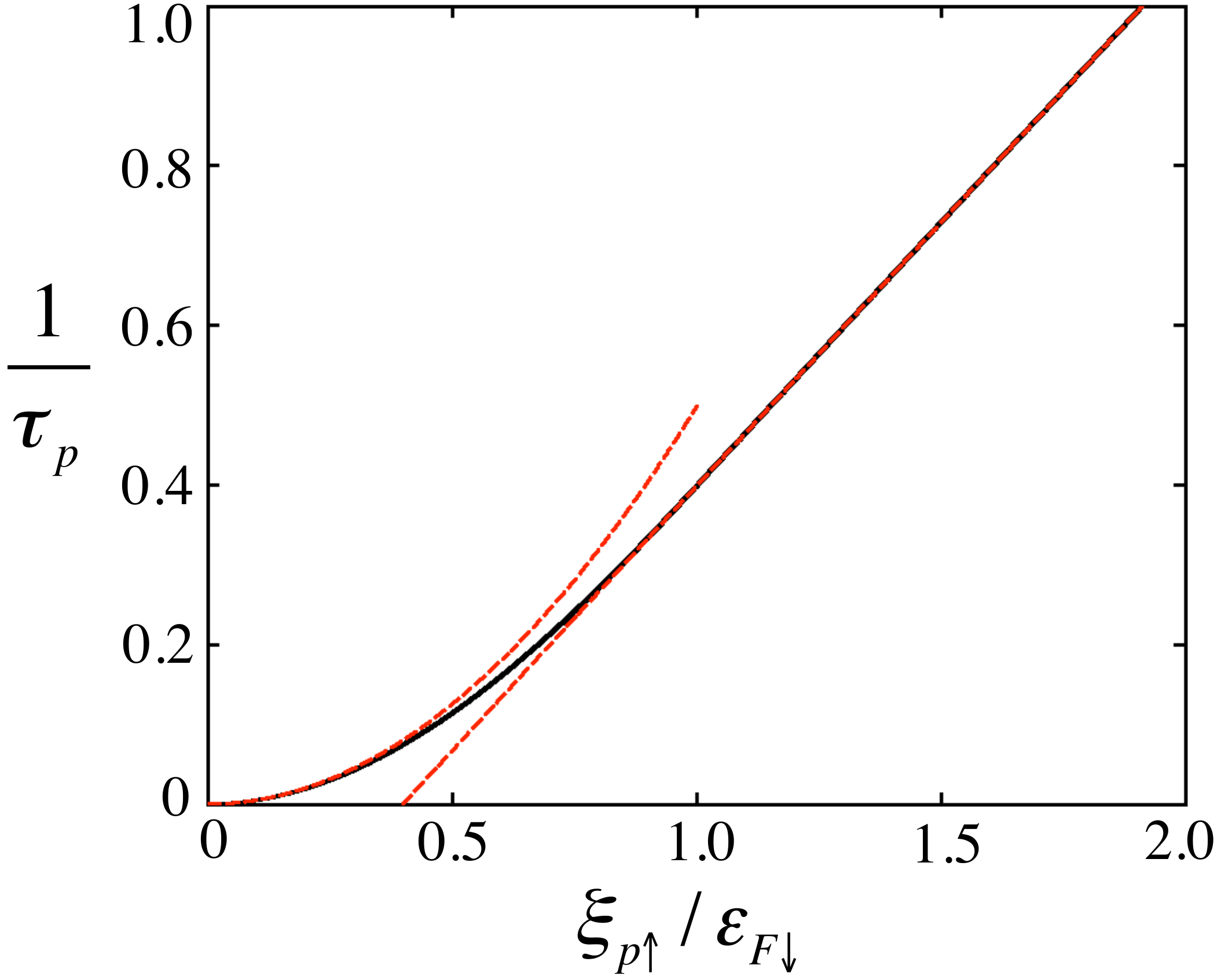}
\caption{(color online). Zero temperature decay rate  $1/\tau_p$ 
(units of $|U|^2m_{\uparrow} m_\downarrow^2\epsilon_{F\downarrow}^2/8\pi^3$)
 of an $\uparrow$ quasiparticles as a function of excitation energy $\xi_{p\uparrow}$, 
with $m_{\downarrow}/m_{\uparrow}=173/6$  and $k_{F\uparrow}/k_{F\downarrow}=2$.
 The  black solid curve  is the result of a numerical integration of (\ref{ImSig}) and the red dashed curves are given by (\ref{exc1}) and (\ref{exc2}). 
 }
\label{exc}
\end{figure}

The linear dependence of  the decay rate  is also characteristic of marginal  Fermi liquids \cite{marginal_FL} although the physics there is quite different and the linear scaling is due to strong spin fluctuations.
In Fig. (\ref{exc}) we plot the decay at zero temperature as a function of the excitation energy. We have chosen the parameters 
$m_{\downarrow}/m_{\uparrow}=173/6$ corresponding to a mixture of $^{173}$Yb and $^6$Li atoms~\cite{Fermi_Fermi3}, and $k_{F\uparrow}/k_{F\downarrow}=2$.
These parameters give $\epsilon_{F\uparrow}/\epsilon_{F\downarrow}=115$ corresponding to a large regime of intermediate energies. Both the usual quadratic and linear scalings are clearly visible in Fig. (\ref{exc}).

{\it Non-zero temperature, degenerate case}.--- In this regime, we have  $0<T\ll T_{F\uparrow}$
and we consider the case  $\xi_{p\uparrow}=0$. From the thermal distribution functions in (\ref{ImSig}) it follows that 
the integrand is significant only in the range $|\omega|\lesssim T$. Since $T\ll T_F^l$, 
we can approximate  (\ref{ImSig}) by
 \begin{equation}
\frac{1}{\tau_{k_{F\uparrow}}}=-\frac{|U|^2m_\uparrow}{2\pi^2k_{F\uparrow}}
\int_0^{2k_{F\uparrow}} qdq\int_{-\infty}^{\infty}
\frac{2{\rm Im}\chi_\downarrow(q,\omega)d\omega}{(e^{\beta\omega}+1)(1-e^{-\beta\omega})}. \label{nonzeroT}
\end{equation}

(i) $T \ll T_{F\downarrow} $.
In this case, we can use the zero temperature, low-energy expression $\text{Im}\chi_{\downarrow}(q,\omega)=-m_{\downarrow}^2\omega/4\pi q$ to obtain
\begin{equation}
\frac{1}{\tau_{k_{F\uparrow}}}=\frac{|U|^2 m_{\uparrow} m_{\downarrow}^2 k_{F\downarrow}} {8\pi k_{F\uparrow}} T^2.
\label{tem1}
\end{equation}
Here the system has the quadratic Fermi liquid behaviour and we recover the standard
result when $T_{F\downarrow}=T_{F\uparrow}$~\cite{book_vignale}.

(ii) $T_{F\downarrow} \ll  T \ll T_{F\uparrow}$. The gas of $\downarrow $ atoms is now classical, and we can use the expression 
$\text{Im}\chi_{\downarrow}(q,\omega, T)=-\pi n_{\downarrow}\sqrt{2 m_{\downarrow}\beta/\pi}\, e^{-\omega^2m_{\downarrow}\beta/2q^2}
e^{-q^2\beta/8m_{\downarrow}}\sinh( \beta\omega/2)/q$~\cite{bruun}. 
The decay rate can then  be written as 
\begin{equation}
\frac{1}{\tau_{k_{F\uparrow}}}=
\frac{2|U|^2m_\uparrow m_\downarrow n_\downarrow T}{\pi^{3/2}k_{F\uparrow}}I(\sqrt{k_{F\uparrow}^2/2m_\downarrow T} )
\label{ImEIntermediateT}
\end{equation}
with 
 \begin{equation}
I(t) \equiv \int _0^tdy \, e^{-y^2}\int_{-\infty}^\infty dx\frac {e^{-x^2/4y^2}}{\cosh x}.
\label{Integral}
\end{equation}
We now discuss two important special cases of (\ref{ImEIntermediateT})-(\ref{Integral}). 
\begin{figure}[!hbp]
\centering
\includegraphics[width=1\columnwidth]{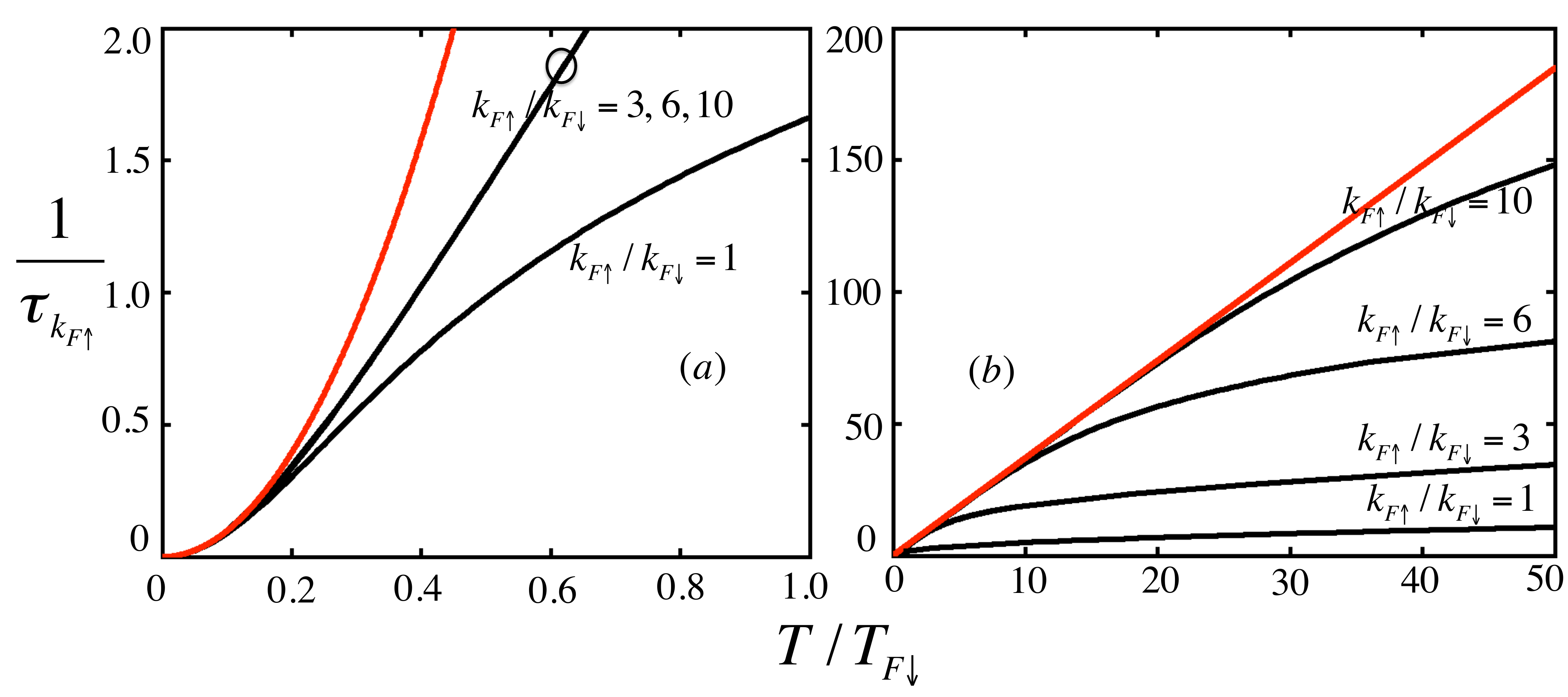}
\caption{(color online) decay rate $1/\tau_{k_{F\uparrow}}$ in units of $|U|^2m^3\epsilon_{F\downarrow}^2k_{F\downarrow}/8\pi^3 k_{F\uparrow}$  as a function of temperature with $m_\uparrow=m_\downarrow=m$ for different density imbalances $k_{F\uparrow}/k_{F\downarrow}=1, 3, 6, 10$. The black (dark) solid  curves  are the numerical integration of  (\ref{ImSig}) while the red (light) solid  curve in (a) and (b) is given by (\ref{tem1}) and (\ref{ImEIntermediateTb}) respectively. }
\label{denimb}
\end{figure}
First we consider the case of a highly polarized system of two spin states of the same atom, i.e.\ $m_\uparrow=m_\uparrow=m$ and 
 $n_{\downarrow} \ll n_{\uparrow}$. 
In this case, we have $ \sqrt{k_{F\uparrow}^2/2mT} =\sqrt{T_{F\uparrow}/T}\gg 1$. Using   
$I(\infty)=\sqrt{ \pi} \ln 2$, we obtain
 \begin{equation}
 \frac{1}{\tau_{k_{F\uparrow}}}=2\ln 2\frac{|U|^2m^2 n_{\rm \downarrow}}{\pi k_{F\uparrow}}T.
\label{ImEIntermediateTb}
\end{equation}
In Fig. \ref{denimb}, we 
plot the decay rate $1/\tau_{k_{F\uparrow}}$ as a function of $T$ for  several density imbalances.
In  Fig. \ref{denimb} (a) we see the quadratic low temperature behaviour. Note that the curves for $k_{F\uparrow}/k_{F\downarrow}=3, 6, 10$ overlap in this range. 
 In  Fig. \ref{denimb} (b) we see the appearance of linear scaling.
 The range of linear temperature scaling increases with the density imbalance.

\begin{figure}[!hbp]
\centering
\includegraphics[width=1\columnwidth]{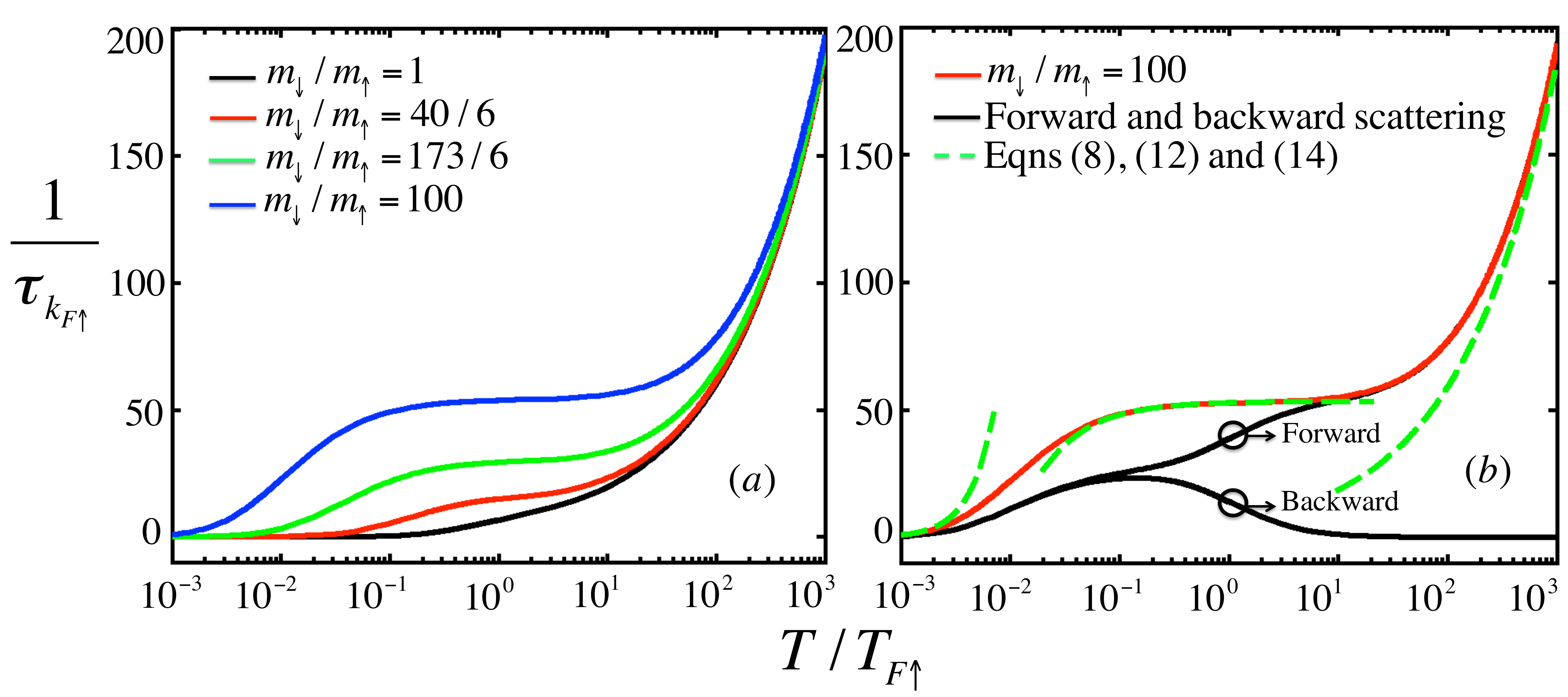}
\caption{(color online) decay rate $1/\tau_{k_{F\uparrow}}$  in units of  $|U|^2k_F^4m_r^2/32\pi^3(m_{\uparrow}m_{\downarrow})^{1/2}$ as a function of temperature for $n_\uparrow=n_\downarrow$  obtained from a numerical integration of (\ref{ImSig}). The left figure shows the  appearance of a plateau 
for $m_\downarrow\gg m_\uparrow$. The right figure compares the numerical results (solid  curves) with (\ref{tem1}),  (\ref{tem2}), and  (\ref{tem3}) (dashed  curves) in the low, intermediate and high temperature limits respectively.
It also shows the contributions of the forward and backward scattering terms of  (\ref{lmSigma}).}
\label{massimb}
\end{figure}

The second case we consider is that  of an equal density mixture of heavy and light atoms ($m_\uparrow\ll m_\downarrow$ 
and $n \equiv n_\uparrow=n_\downarrow$) so that $ \sqrt{k_{F}^2/2m_\downarrow T} =\sqrt{T_{F\downarrow}/T}\ll 1$. Since $I(t) \simeq \sqrt\pi t^2(1-t^2)$
 for $t\ll 1$ we find that 
  \begin{equation} 
 \frac{1}{\tau_{k_{F\uparrow}}}=\frac{ |U|^2 m_{\uparrow}  n_{\downarrow} k_{F\uparrow} } {\pi}(1-\frac{T_{F\downarrow}}{T})
\label{tem2}
\end{equation}
 which shows that the decay rate is constant to leading order in $T_{F\downarrow}/T$. 
This  peculiar behaviour can be understood as follows. 
 Since $T \ll T_{F\uparrow}$, we have $v_{\rm th\downarrow}\ll v_{F\uparrow}$ with $v_{\rm th\downarrow}=\sqrt{2T/m_\downarrow}$
  and  $v_{F\uparrow}= k_{F\uparrow}/m_\uparrow$, and  the $\downarrow$  atoms are moving very slowly compared with the $\uparrow$ atoms. The 
 decay rate is then dominated by the motion of the $\uparrow$ quasiparticle: $\tau^{-1}= n_{\downarrow}\sigma_{\rm sc} v_{F\uparrow}$ where 
  where $\sigma_{\rm sc}$ is the scattering cross section. Using $\sigma_{\rm sc}=|U|^2 m_{\uparrow}^2/\pi$ for $m_\downarrow\gg m_\uparrow$~\cite{bruun}, we get
  $\tau^{-1} \sim |U|^2m_\uparrow n_{\downarrow} k_{F\uparrow} /\pi$, which is precisely the leading constant term of the above expression. 
  This constant term in (\ref{tem2}) is well-known in the theory of doped semiconductors or dirty metals since there the heavy atoms correspond to the static impurities \cite{dirty}.
 In  Fig. \ref{massimb}, we plot the decay rate of an $\uparrow$ excitation with momentum $k_{F\uparrow}$ as a function of temperature. 
 The emergence with increasing mass ratio $m_\downarrow/m_\uparrow$ of a plateau where the rate is independent of temperature 
 is clearly visible in Fig. \ref{massimb} (a). Note that this plateau extends to temperatures well above $T_{F\uparrow}$
 for large mass ratios even though the simple kinetic argument breaks down above $T_{F\uparrow}$ since we then have to take into account the changes to the $\uparrow$ Fermi surface which also affect the scattering rate. Instead, as we show below, the reason the plateau continues to higher temperatures is because of a cancellation between the changes to the forward and backward scattering terms.

{\it Non-zero temperature, classical case ($p=k_{F\uparrow}$)}.--- In the regime $T_{F\uparrow} \ll T$, both  $\sigma=\uparrow,\downarrow$ distributions    are classical.

(i) $ T_{F\uparrow} \ll T\ll T_{F\uparrow} m_{\downarrow}/m_{\uparrow}$. Assuming that $U$ continues to be  weakly dependent on energy and momentum (as occurs for example when $k_{F\uparrow}|a| \ll1$),  the plateau  will persist to a much larger temperature scale, given by $v_{\rm th \downarrow}\sim v_{F\uparrow}$ (i.e. $T\sim T_{F\uparrow} m_{\downarrow}/m_{\uparrow}$) as can be seen from the numerics (see Fig. (\ref{massimb})). This seems to indicate that the Pauli blocking of the $\uparrow$ atoms plays no role for the decay rate of an $\uparrow$ quasiparticle since the plateau survives independently of whether the $\uparrow$ atoms are degenerate ($T \ll T_{F\uparrow} $) or classical ($T \gg T_{F\uparrow} $). To better understand this unusual behavior, we plot in Fig. \ref{massimb} (b),  the forward and backward scattering contributions to $1/\tau_{k_{F\uparrow}}$ with $m_{\downarrow}/m_{\uparrow}=100$ \cite{opticallattice}. We find that, when $T \ll T_{F\uparrow} $, the forward and backward scatterings contribute equally to $1/\tau_{k_{F\uparrow}}$, which can be understood by taking $\xi_{\bf p\uparrow}=0$ in (\ref{ImSig}).  In the regime $T_F{\uparrow} \lesssim T \ll T_{F\uparrow}(m_{\downarrow}/m_{\uparrow})$,  the backward scattering begins to decrease while the forward scattering increases, keeping however their sum constant. This shows that Pauli blocking indeed affects the forward and backward scatterings but not the sum of the two. This constant sum is due to the vanishing energy transfer in the decay process of the $\uparrow$ quasiparticle with a large mass imbalance ( $ \omega_{max}/T\sim \max(m_{\uparrow}/m_{\downarrow}, \sqrt{T_{F\downarrow}/T})\ll 1)$. Thus by rewriting $n_{\bf k\downarrow } (1-n_{ {\bf k+q}\downarrow} ) (1-n_{{\bf p-q}\uparrow } )+(1-n_{ \bf k\downarrow} )n_{{\bf k+q}\downarrow }=(n_{\bf k\downarrow }-n_{ {\bf k+q}\downarrow})/(e^{\beta\omega}-1)[e^{\beta\omega}(1-n_{{\bf p-q}\uparrow })+n_{{\bf p-q}\uparrow }]\simeq (n_{\bf k\downarrow }-n_{ {\bf k+q}\downarrow})/(e^{\beta\omega}-1)$, we see that the $\uparrow$ atoms play no role in the integrand of (\ref{ImSig}),  explaining why the plateau survives independently of whether the $\uparrow$ atoms are degenerate or classical, i.e., the Pauli blocking is irrelevant in this case. This regime occurs also in warm dense plasmas where the electrons are close to degeneracy but the ions are already classical \cite{wdm}. Note however that the lifetime is likely to behave very differently since the electron-ion cross section is momentum dependent, unlike the low-energy contact interaction.

(ii) $ T_{F\uparrow} m_{\downarrow}/m_{\uparrow}\ll T$. Using the classical limit of the chemical potential for fixed particle number in  (\ref{ImSig}), i.e.,  
$\mu/T\rightarrow-\infty$ when $T\rightarrow \infty$, we find $1/\tau_{k_{F\uparrow}}$ is again given by  (\ref{ImEIntermediateT}), but with
\begin{equation}
I(t)=\int_0^{\infty}  dy e^{-y^2} \int_{(-y^2-y t)\frac{2m_\downarrow}{m_\uparrow}}^{(-y^2+y t)\frac{2m_\downarrow}{m_\uparrow}}dx e^{-x^2/4y^2} e^{x}.
\label{Integral2}
\end{equation}
where $t=\sqrt{k_{F\uparrow}^2/2m_\downarrow T}$.
When $t  \ll1$ the integral can be evaluated  straightforwardly, which gives $I(t)=2t m_r^2/m_\uparrow m_\downarrow$, and the decay rate becomes
\begin{equation} 
\frac{1}{\tau_{k_{F\uparrow}}} =\frac{ 2\sqrt{2}|U|^2m_r^2   n_{\downarrow}  } { \pi^{3/2}\sqrt{m_\downarrow}} \sqrt{T}.
\label{tem3}
\end{equation}
It is interesting to compare (\ref{tem3}) with (\ref{tem2}). First we note that the Fermi momentum $k_{F\uparrow}$ in (\ref{tem2})
has been replaced by $\sqrt T$ in  (\ref{tem3})  as expected for the high temperature regime. Second, when the two expressions 
are equated we obtain that the  cross-over between the intermediate $T$ behaviour given by (\ref{tem2}) and the high $T$ behaviour given by 
 (\ref{tem3})   occurs for $T\sim T_{F\uparrow}m_\downarrow/m_\uparrow$ in agreement with the analysis above. Thus, the region where the damping rate is independent 
of temperature when $n_\uparrow=n_\downarrow$ and $m_\downarrow\gg m_\uparrow$ extends to temperatures much higher than $T_{F\uparrow}$. 
This effect is clearly illustrated in Fig. (\ref{massimb}). Importantly, it makes the experimental observation of this plateau regime significantly easier
as it emerges already for fairly high temperatures, when the mass imbalance is large.

{\it Polaron case}.--- Given its experimental importance, we finally calculate the collision rate of the minority particles.
 When   $n_\downarrow \rightarrow 0$, i.e., in the polaron limit, we find from  (\ref{lifetime})
\begin{equation}
\frac1{\tau_\downarrow}= \frac{4}{15\pi^3}|U|^2m_\downarrow m_\uparrow^2 \left( \epsilon_{\downarrow}^2+\frac{5\pi^2 T^2 }{32 } \right)
\label{polaron}
\end{equation}
for $\epsilon_{\downarrow}\ll\epsilon_{F\uparrow}$ and $T\ll T_{F\uparrow}$. The polaron therefore 
shows normal Fermi liquid behavior, consistent with the literature~\cite{polaron_theory}.

{\it Experimental probes}.--- 
Radio-frequency (RF) spectroscopy is a very successful method to probe the single particle properties in cold atom gases, and it is well 
suited to detect these new scaling laws. First we note that the intermediate temperature regime
 $T_{F\downarrow}\ll T\ll T_{F\uparrow}$ has already been achieved by the Innsbruck group using a mixture of $^6$Li and $^{40}$K 
 atoms~\cite{polaron_rudi}. They  probed the single particle properties of the $\downarrow$  atoms ($^{40}$K) 
 using  RF spectroscopy, where  the $^{40}$K atoms performed Rabi oscillations between an interacting and a non-interacting state. 
 Quasiparticle collisions cause decoherence, and the observed damping of the Rabi oscillations can therefore be used to measure the collision rate.
 A similar experiment  should be able to detect 
 the $\uparrow$ collision rate as described by  (\ref{lifetime}). One could block out the low lying states in the Fermi sea of $\uparrow$ atoms from participating in the RF 
 spectroscopy by coupling them to a non-interacting state with a filled Fermi sea slightly smaller than that of the  $\uparrow$ atoms.
The low lying states are then inert to the RF probe due to the blocking of the non-interacting state. 
Alternatively, one could use  momentum-resolved RF spectroscopies ~\cite{jin, polaron_rudi, polaron_kohl}.  For $^6$Li-$^{40}$K  mixture, the typical experimental parameters used in~\cite{polaron_rudi} are $\epsilon_F^{\text{Li}} =232\, \hbar \, $kHz,  $k_F^{ \text{Li}}=\hbar^{-1}\sqrt{2m_{\text{Li}}\epsilon_F^{\text{Li}}}=1/2850a_0$ with $a_0$ Bohr's radius and the interspecies background scattering length $a_{bg}=63 \, a_0$.  Assuming equal density $n_{\text{Li}}=n_{\text K}$, and $k_F a=0.1$, we find the quasiparticle lifetime of Li in the plateau regime is  $\sim 1$ms, which is the typical scale measured in ~\cite{polaron_rudi}. In solid state systems, deviations from standard quadratic behaviour often show up in electrical resistivity, specific heat and magnetic susceptibility measurements.  In atomic gases, our results could affect  transport properties such as the spin drag rate (which has already been measured \cite{Sommer}), the damping of collective modes, and thermodynamic properties such as the heat capacity.

{\it  Discussion and conclusions}.--- We have shown that weakly interacting Fermi-Fermi mixtures realise a rich diversity of regimes for the quasiparticle damping 
which are analogous to several   quite distinct physical systems in nature. 
These regimes are characterised by scaling laws for the quaisparticle lifetime which are different from the usual quadratic case. Our results are derived in the weakly interacting limit 
making them quantitatively reliable.  The effects described in this letter can be measured using RF spectroscopy.

{\it  Acknowledgements}.--- We would like to thank M. Baranov, P. Massignan and T. Killian for discussions. ZL and CL acknowledge support from the EPSRC through grant EP/I018514/1.


\begin{references}
\bibitem{opticallattice}  Even for equal mass spin mixtures, an unequal mass mixture can be created by e.g. imposing a spin dependent optical lattice which can change the effective mass of one of the species by extremely large values.
\bibitem{note2} In the case of atomic gases we also assume that 3-body recombination rates are small and do not affect the lifetime.
\end{references}
\end{document}